\documentclass[twocolumn]{mn2e}
\usepackage{etoolbox}
\usepackage{epsfig}
\usepackage{epstopdf}
\usepackage{subfloat}
\usepackage{subfigure}
\usepackage{multirow}
\usepackage{graphicx,times}  
\usepackage{soul}
\usepackage{natbib}
\usepackage{amssymb,amsmath}
\usepackage{hyperref}

\def\apjl{APJL}
\def\aap{AAP}

\def\HI{{H~{\sc i} }}

\begin{document}

\title[Strong lensing to probe post reionization \HI]{Using strong gravitational lensing to probe the post reionization \HI power spectrum}
\author[Urvashi Arora and Prasun Dutta ]
{Urvashi Arora$^{1}$\thanks{Email: urvashi.rs.phy17@itbhu.ac.in}, 
Prasun Dutta$^{1}$\thanks{Email:pdutta.phy@itbhu.ac.in},  
\\$^{1}$ Department of Physics, IIT (BHU) Varanasi, 221005  India. 
}
\maketitle 

\begin{abstract}
Probing  statistical distribution of the neutral hydrogen (\HI) using the redshifted 21-cm hyperfine-transition spectral line  holds the key to understand the formation and evolution of  the matter density  in the universe. The two point statistics of the \HI distribution  can be estimated by measuring the power spectrum of the redshifted 21-cm signal using visibility correlation. A major challenge in this regard is that the expected signal is weak compared to the foreground contribution from the Galactic synchrotron emission and extragalactic point sources in the observing frequencies. In this work, we investigate the possibility of detecting the power spectrum of the redshifted 21-cm signal by using strong gravitational lensing of the galaxy clusters. This method has the advantage that it only enhances the \HI signal and not the diffuse galactic foreground. Based on four simple models of the cluster potentials we show that the strong lenses at relatively lower redshifts with more than one dark matter halo significantly enhances the 21-cm signal from the post reionization era.  We discuss the merits and demerits of the method and the future studies required for further investigations.

\end{abstract}


\section{Introduction}
It is now well established that the small perturbations in the initial matter density in the universe grew under gravitational instabilities and eventually formed the first luminous objects \citep{1980lssu.book.....P,1989RvMP...61..185S,
1999coph.book.....P, 2002thas.book.....P,2002PhR...367....1B}. The universe in its earlier stage was predominantly filled with \HI, which got ionised by the radiation from the first luminous objects during the epoch of reionization. Observational evidence suggests that by the redshift of $\sim 6$ most of the universe is completely ionised with traces of \HI left in places with baryonic overdensities (see for example, \citet{2006ARA&A..44..415F} and the references therein). These are also the places where the structures in the post reionization universe grew and the present galaxies and galaxy clusters originated. In this work, we focus our attention on probing the structure formation in the post reionization universe.

\citet{1972A&A....20..189S, 1975MNRAS.171..375S}  suggested using the redshifted 21-cm hyperfine structure transition of \HI to study the distribution of baryonic matter in the post reionization era. \citet{2001JApA...22...21B, 2009MNRAS.397.1926W} uses the power spectrum of the redshifted 21-cm intensity fluctuation to probe the statistical distribution of \HI.  \citet{2001JApA...22...21B, 2005MNRAS.356.1519B} develops a methodology to estimate this \HI power spectrum by using the radio interferometric observations. They show that correlating the directly observed quantity visibility by the interferometers in nearby baselines, it is possible to measure the \HI power spectra unbiasedly.  Several efforts are made to model the \HI  power spectra using analytic, semi analytic models and simulations \citep{2004JApA...25...67B, 2009MNRAS.397.1926W, 2010MNRAS.407..567B, 2014MNRAS.444.3183A} including the \HI bias \citep{2012MNRAS.421.3570G, 2016MNRAS.460.4310S, 2017MNRAS.471.1788C, 2018MNRAS.474.5143M, 2018MNRAS.476...96S}  and the redshift space distortion effects \citep{2012MNRAS.425.2128J, 2016MNRAS.455L..77H, 2019MNRAS.487.5666S, 2019JCAP...09..024M}. These studies collectively provide a fair description of the expected redshifted 21-cm signal from the post reionization era.

All these models predict a fairly weak amplitude for the visibility correlation signal of less than $1$ mJy$^{2}$ in the detection range of the existing telescopes like the GMRT \footnote{GMRT: Giant Meterwave Radio Telescope, Pune, India}, the MWA \footnote{MWA: Murchison Widefield Array, Australia} or the LOFAR \footnote{LOFAR: Low-Frequency Array, Netherlands}  \citep{2010MNRAS.407..567B}. Observationally, one aims to detect these signals by choosing a radio-quiet part of the sky. However, contribution from the diffuse  Galactic synchrotron emission and extra galactic compact sources add up to the redshifted 21-cm signal in the observing frequencies. These are usually referred to as foreground contamination to the signal and have  several orders of magnitude higher amplitude \citep{2016ApJ...818..139T} compared to the signal itself. A majority of work \citep{2010ApJ...724..526D, 2011MNRAS.418.2584G, 2016MNRAS.463.4093C} over the last decade towards the  \HI power spectrum estimation  focuses on better detection, mitigation and avoidance of these foreground contaminations \citep{ 2017NewA...57...94C,    2019MNRAS.490..243C, 2019MNRAS.487.4102C,
2020MNRAS.494.3392C}.

Galaxy clusters act as  spatially distributed strong gravitational lenses (see \citet{2011A&ARv..19...47K} for a review) and hence may be used to detect the spatial statistics of the \HI distribution at higher redshifts than the cluster. The lensing mass distributions in a galaxy cluster can be modelled using multiple optical  images of lensed galaxies \citep{2010MNRAS.402L..44R, 2014MNRAS.444..268R, 2015MNRAS.452.1437J, 2016A&A...588A..99L, 2016MNRAS.457.2029J, 2018ApJ...859..159C}. Such models are then often used to  study the spatially resolved properties  of high redshift galaxies (for example see \citet{2018MNRAS.481.1427S}).  
\citet{2001ApJ...557..421S}  investigated the possibility to study post reionization \HI distribution by observing individual galaxies through strong lenses. \citet{2015MNRAS.452L..49D}  estimate the possibility of  detecting \HI emission from strongly lensed individual galaxies using existing and future radio interferometers. The possibility of detecting post reionization \HI statistics using weak gravitational lensing is discussed in \citet{2015MNRAS.448.2368P}.  \citet{2019MNRAS.484.3681B} gives the first observational effort to detect individual strongly lensed galaxies using the GMRT. 

In this work, we investigate the possibility of probing the redshifted 21-cm power spectrum using its enhancement by strong gravitational lensing of galaxy clusters. We analytically calculate the modification in the visibility correlation signal in the presence of strong lensing. To access this effect in terms of observations, we consider four simple lensing models. We use a detailed model of post reionization \HI bias and redshift space distortion effects along with these simple cluster models to estimate the enhanced visibility correlation signals for two fiducial redshifts of the lensing clusters and three fiducial redshifts of the source \HI distribution. The rest of the paper is arranged in the following way. In section 2 we provide the analytical formalism to calculate the  modified  visibility correlation. The expected lensed visibility correlation signal from our four simple lens models are shown in section 3. We discuss the merits and demerits of this technique in section 4 and conclude.

\section{Formalism}
\subsection{Visibility function of lensed \HI emission}
Specific intensity of \HI emission at the observer $ I_{obs}(\vec{\theta}, z_{s}, z_{L})$, originated from a direction $\vec{\theta}$ in the sky (with respect to the centre of the field of view of observation) at the redshift $z_{s}$ and modified by a strong gravitational lens at the redshift $z_{L}$ can be written as
\begin{equation}
I_{obs}(\vec{\theta}, z_{s}, z_{L})= \int d \vec{\theta'} \left [  I_{0} + \delta I(\vec{\theta'}, z_{s}) \right ]\, G_{L}(\vec{\theta}-\vec{\theta'}, z_{s}, z_{L} ),
\end{equation}
where $G_L(\vec{\theta}, z_s, z_L)$ is the point spread function  of the gravitational lens. Redshift dependence on the specific intensity at the observer arises  from (i) the redshifted frequency of the 21-cm emission and (ii) the redshift dependence of the point spread function of the gravitational lens $G_L$. The point spread function of the gravitational lens also depends on the angular direction in the sky. We shall discuss these dependencies in detail in a later section. The observed visibility by radio interferometers of such lensed \HI emission can be written as
\begin{eqnarray}
V(\vec{U}, z_{s}, z_{L} )=&\int d\vec{\theta'} \int d\vec{\theta}\, \delta I(\vec{\theta'}, z_{s})\,A(\vec{\theta}, z_{s}) \, \nonumber  \\
&G_{L}(\vec{\theta}-\vec{\theta'}, z_{s}, z_{L} )\exp({-i2\pi \vec{U}.\vec{\theta}}). 
\end{eqnarray}
We have assumed here that the angular extent of the gravitational lens is small and the flat sky approximation is valid.
The interferometers usually do not observe at zero baselines, hence in absence of lensing, the first term in the integral of eqn.~(1) is not measured. Strong lensing, however, modifies the first term by the point spread function of the lens $G_L$. However, since lensing conserves the surface brightness, contribution through the first term in  eqn.~(1) in visibility remains the same and is not measured by the interferometers. The antenna beam pattern $A(\vec{\theta})$ of the telescope depends on the source redshift through the observing frequency and  is related to the aperture function $\tilde{a}(\vec{U})$ as 
\begin{equation}
A(\vec{\theta})=\int d\vec{U}\,  \tilde{a}(\vec{U})\,  e^{i2 \pi \vec{U}. \vec{\theta}}.
\end{equation}
In this work, we investigate the effect of strong lensing on intensity mapping of the redshifted 21-cm emission from the neutral hydrogen. 21-cm intensity mapping is discussed in \citet{2001JApA...22...21B} and \citet{2005MNRAS.356.1519B}.  \citet{2005MNRAS.356.1519B} define the 21-cm radiation efficiency $\eta_{HI}(\vec{x})$ as
\begin{equation}
\eta_{HI}(\vec{x}, z_s) = \frac{\rho_{HI}(\vec{x}, z_s)}{\bar{\rho}_H(z_s)} \left ( 1 - \frac{T_{\gamma} (z_s)}{T_s (z_s)} \right ) \left [ 1 - \frac{(1+z_s) }{H(z_s)} \frac{\partial v}{\partial r} \right ],
\end{equation}
where $\rho_{HI}(\vec{x}, z_s) $ is the density of \HI at the point $\vec{x}$ in the source redshift, 
$\bar{\rho}_H(z_s)$ is the mean hydrogen density at $z_s$,  $T_{\gamma}$ and $T_s$ are the mean CMB and spin temperature at the source redshift  and $v$ give the peculiar velocity.
The brightness temperature fluctuation in 21-cm emission is related to the 21-cm radiation efficiency $\eta_{HI}(\vec{x})$ as 
\begin{equation}
\delta I(\vec{\theta}, z_{s})=\frac{\partial B}{\partial T} \bar{T}(z_{s}) \eta_{\text{\HI}}(\vec{x}, z_{s}),
\end{equation}
where $B$ is the Plank's function and  $\bar{T}(z_{s})$ can be calculated for a background cosmology as 
\begin{equation}
\bar{T}(z_s)=4.0 mK(1+z_s)^2\left(\frac{\Omega_b h^2}{0.02}\right)\left(\frac{0.7}{h}\right)\left(\frac{H_0}{H(z_s)}\right).
\end{equation}
We define the lensing sampling function $S_{L}(\vec{U}, z_s, z_L)$  as
\begin{equation}
G_{L}(\vec{\theta}, z_{s}, z_{L})= \int d\vec{U} S_L(\vec{U}, z_{s}, z_{L})\, e^{i 2 \pi \vec{U}. \vec{\theta}}.
\end{equation}
The observed (lensed) visibility can be written as
\begin{eqnarray}
 V(\vec{U}', z_{s}, z_{L}) = {\bar{T}(z_{s})}\ \frac{\partial B}{\partial T}  \int \frac{d\vec{k}}{(2 \pi)^3}  \eta_{\text{\HI}}(\vec{k})\, \nonumber \\   S_L \left(\frac{r_\nu \vec{k}_\perp}{2 \pi}, z_{s}, z_{L}\right) \ \tilde{a}\left(\vec{U}-\frac{r_\nu \vec{k}_\perp}{2 \pi}\right) e^{i k_\parallel r_\nu}, 
 \end{eqnarray} 
where the quantity $\tilde{\eta}_{\text{\HI}}(\vec{k}, z_{s})$ gives the 21-cm radiation efficiency in the Fourier space as 
\begin{equation}
\eta_{HI}(\vec{x}, z_{s})=\int \frac{d\vec{k}}{(2\pi)^3}\,  \tilde{\eta}_{\text{\HI}}(\vec{k}, z_{s})\, e^{i\vec{k}.\vec{x}}.
\end{equation}
We have used $\vec{k}.\vec{x}=r_\nu[k_\parallel + \vec{K}_\perp.\vec{\theta}]$ with $\parallel$ denoting the component of the vector $\vec{k}$ along the line of sight of observation and $\perp$ the components in the plane of the sky, $\vec{x}$, the comoving position vector. The quantity $r_{\nu}$ denotes the comoving distance to the redshift, where for the observing frequency $\nu$ and rest frequency of 21-cm emission $\nu_0$,  $z_s = \nu_0/\nu - 1 $.
 
\subsection{Power spectrum for the lensed \HI emission}
 We define the \HI power spectrum at any redshift as
\begin{equation}
<\eta^*_{\text{\HI}}(\vec{k}) \, \eta_{\text{\HI}}(\vec{k}')>=(2  \pi)^3  \delta(\vec{k}-\vec{k}') P_{\text{\HI}}(\vec{k}').
\end{equation}
 \citet{2001JApA...22...21B} and \citet{2005MNRAS.356.1519B}  show that the visibility correlation at nearby baselines $V_2(\vec{U}, \vec{U'}, \nu, \Delta \nu) = \langle V^*(\vec{U},\nu)V(\vec{U'},\nu + \Delta \nu) \rangle $ has the information of the redshifted 21-cm power spectrum. Here we estimate the observed visibility correlation for a given 21-cm power spectrum at redshift $z_s$ modified by an intervening strong gravitational lens at redshift $z_l$. We denote the visibility correlation with lensing as $V_{2L}$, where
\begin{eqnarray}
\nonumber & V_{2L}(\vec{U}, \vec{U'}, \nu, \Delta \nu)  =  C'(z) \int \frac{d\vec{k_{\parallel}} d\vec{k_{\perp}}}{(2 \pi)^3}   
 e^{-i k_\parallel r'_\nu \Delta \nu}\\ \nonumber & S_L^{*}\left(\frac{r_\nu \vec{k}_\perp}{2 \pi}, z_{s}, z_{L} \right) S_L\left(\frac{r_{\nu_1} \vec{k}_\perp}{2 \pi}, z_{s}, z_{L} \right)  \tilde{a}^{*}(\vec{U}-\frac{r_\nu \vec{k}_\perp}{2 \pi}) \\ & \tilde{a}\left(\vec{U}'-\frac{r_{\nu_1} \vec{k}_\perp}{2 \pi}\right) P_{\text{\HI}}(\vec{k}).  
\end{eqnarray}
Here $\nu_1 = \nu + \Delta \nu$, $C'(z)= \left [ \bar{T}(z_s)  \frac{\partial B}{\partial T} \right ]^2$, $r_{\nu_1}$ is the comoving distance at frequency $\nu+\Delta \nu$
\begin{equation}
r_{\nu_1}=r(\nu)+  \frac{\partial r_\nu}{\partial \nu} \Delta \nu=r_{\nu}+ r_\nu' \Delta\nu.  
\end{equation} 
The aperture function of a telescope depends on the dipole pattern as well as the antenna dimensions. 
For the purpose of this calculation, we assume  $A(\vec{\theta})=\exp(-\frac{\theta^2}{\theta_0^2})$, where the parameter $\theta_0$ quantifies the field of view of the observation. The functions $S_L$ and $\tilde{a}$ depends weekly on $r_\nu$, where we assume $S_L( r_{\nu_1} \vec{k}_\perp/ 2 \pi , z_{s}, z_{L}) = S_L ( r_{\nu} \vec{k}_\perp / 2 \pi , z_{s}, z_{L}) $ and  $ \tilde{a}(\vec{U}'- r_{\nu_1} \vec{k}_\perp / 2 \pi) = \tilde{a} (\vec{U}'- r_{\nu} \vec{k}_\perp / 2 \pi )$. For  $ k_{\perp} >>\frac{2}{r_\nu \theta_0}$ the visibility correlation can be written as
\begin{eqnarray}
 V_{2L}&(\vec{U},\vec{U}',\nu,\Delta \nu) = C(z) \mid S_L(\vec{U}, z_{s}, z_{L}) \mid ^2 \int dk_\parallel  e^{-i k_\parallel r_\nu' \Delta \nu} \nonumber  \\
 & \times    \exp \left(-\frac{|\frac{2 \pi}{r_\nu}(\vec{U}-\vec{U}')|^2}{(\frac{2}{r_\nu \theta_0})^2}\right) P_{\text{\HI}}(\sqrt{k_\parallel^2+(\frac{2 \pi \vec{U}}{r_\nu})^2}),
 \label{eq:vL}
\end{eqnarray}
where $C(z) = C'(z) \theta_0^2 / 2 r_\nu^2$. 

\citet{2001JApA...22...21B} considered the visibility correlation at nearby baselines to estimate the \HI power spectrum from the observed visibilities. The nearby baseline correlation reduces the noise bias that appears by correlating the visibilities in the same baselines and  introduces the factor $ \exp \left(-\frac{|\frac{2 \pi}{r_\nu}(\vec{U}-\vec{U}'|^2}{(\frac{2}{r_\nu \theta_0})^2}\right)$. In this work, we do not consider the effect of measurement noise in the visibilities and correlate the visibilities at the same baselines. Furthermore, the visibility correlation is also done at the same frequencies. Hence eqn~(\ref{eq:vL}) simplifies to
\begin{eqnarray}
V_{2L}(\vec{U},\nu)=C(z)  | S_L(\vec{U}, z_{s}, z_{L})|^2 \int dk_\parallel    P_{\text{\HI}}( \sqrt {k_\parallel^2+(\frac{2 \pi \vec{U}}{r_\nu})^2}).
\label{eq:V2L}
\end{eqnarray}
Clearly, the effect of gravitational lensing is a magnification of the visibility correlation by a factor of $\mid S_L \mid^2$. 

\section{Simulating lensed 21-cm power spectrum}

In this section, we discuss the methodology to simulate lensed 21-cm power spectrum from the post reionization universe. We first model the 21-cm power spectrum  from the expected dark matter distribution at post reionization redshifts, a  scale-dependent \HI bias and redshift space distortion effects. We use four simple models for the lensing potential based on strong lensing by galaxy clusters and calculate the corresponding lensing sampling functions.

\subsection{Modelling redshifted 21-cm power spectrum}
\begin{figure}[H]
   \vspace{-0.5cm}
    \hspace{-0.7cm}
    \includegraphics[width=9cm,height=6cm]{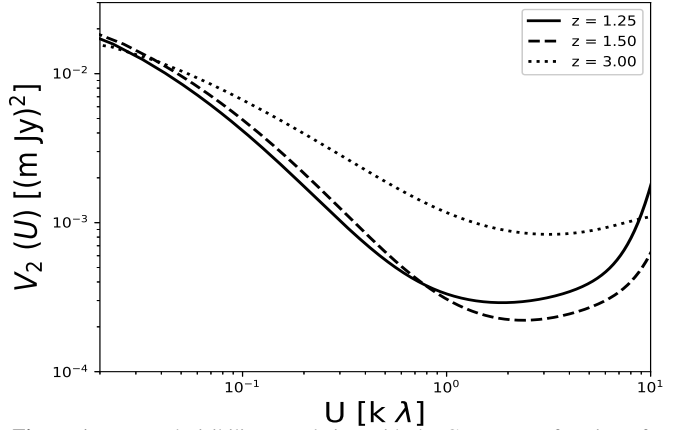}
    \caption{Expected visibility correlation with the GMRT as a function of baselines for three post reionization redshifts. }
    \label{fig:v2}
\end{figure}
Model for the  \HI power spectrum in the post reionization era  is discussed in \citet{2005MNRAS.356.1519B} where they use the dark matter power spectrum and a scale independent bias to model the \HI signal. Better models for the signal are presented is \citet{2016MNRAS.460.4310S,2018MNRAS.476...96S}, where they use numerical simulation to calculate  the effect of scale dependent bias as well as redshift space distortion factor.  The \HI power spectrum $P_{HI}(\vec{k_{\perp}}, k_{\parallel})$ can be written in terms of the matter power spectrum $P(k)$ as
\begin{eqnarray}
P_{HI}(\vec{k_{\perp}}, k_{\parallel}) = & b(k)^2 \ \left [ 1 + 2 r \beta  \mu^2 + \beta^2 \mu^4 \right ] \nonumber \\  & D_{FoG} (k_{\parallel}, \sigma_p) P(k), 
\label{eq:HIPS}
\end{eqnarray}
where $b$ is the scale dependent bias and  $\sigma_p$ is a parameter for the redshifted space distortion. Both of these depends on the redshift and wave number. The factor $\mu = k_{\parallel}/k$.  \citet{2016MNRAS.460.4310S,2018MNRAS.476...96S}  show  that the bias is a complex function where $r$ gives the ratio of its real component to the amplitude. They provide fitting functions to their estimated bias and redshift space distortion over a comoving wavenumber range of $0.01 - 10.0$ Mpc$^{-1}$  and redshift range of $1$ to $6$.  We use the dark matter power spectrum from \citet{1994MNRAS.267.1020P} along with a standard $\Lambda$CDM cosmological parameters from \citet{2016A&A...594A..13P} . 

Figure~\ref{fig:v2} shows the visibility correlation signal as expected for observation with the GMRT  at three post reionization redshifts based on the  \HI power spectrum given in eqn.~(\ref{eq:HIPS}). It is known that  in $\Lambda$CDM cosmology, the angular diameter distance to any object at a redshift $z$ increases to a redshift of $\sim 1.5$ and show a slow decrease with redshift afterwards. The luminosity distance to an object, on the other hand, increases monotonically with redshift. This makes any high redshift object appear increasingly smaller in angular size with decreasing flux as the redshift changes upto a redshift of $1.5$.  For astrophysical objects at higher redshifts, the flux decreases continuously, however, the angular size shows a slight  increase with redshift. We choose three redshifts for this study, redshift of $1.25, 1.5$ and $3.0$, to scan the effects of the above change in redshifts. The comoving wavenumber range of $0.2 - 10.0$ Mpc$^{-1}$ corresponds to a baseline range of $\sim 200\ \lambda$  to $\sim 10 \, {\rm k}\lambda$ at a redshift of $3.0$.   At the redshift of $1.25$, the baseline  corresponding to the comoving wavenumber  of $0.2$ Mpc$^{-1}$ is  $125 \lambda$ and a wavenumber of $10.0$ Mpc$^{-1}$ gives a baseline of
$\sim 6 \, {\rm k}\lambda$.  At the lower end of the wave numbers, $\sim 0.2$ Mpc$^{-1}$,  the \HI power spectra mostly traces the dark matter power spectra modified by a scale dependent bias.  At wave-numbers of $>0.4$ M pc$^{-1}$  the redshift space distortion produced by peculiar velocity becomes important. At the highest $k \sim 10$ Mpc$^{-1}$ the \HI power spectra probes the finger of god effects in individual over-density regions.  Thus, strong lensing may provide a good probe of the scale dependent bias and the redshift space distortion effects and their evolution over redshift. 
 Note that, for this choice of baseline ranges, for some redshifts, the visibility correlations plotted here requires extrapolation from the ranges of wave numbers used in \cite{2018MNRAS.476...96S}.

\subsection{Model for lensing potential}
\begin{figure*}
  \includegraphics[width=\textwidth]{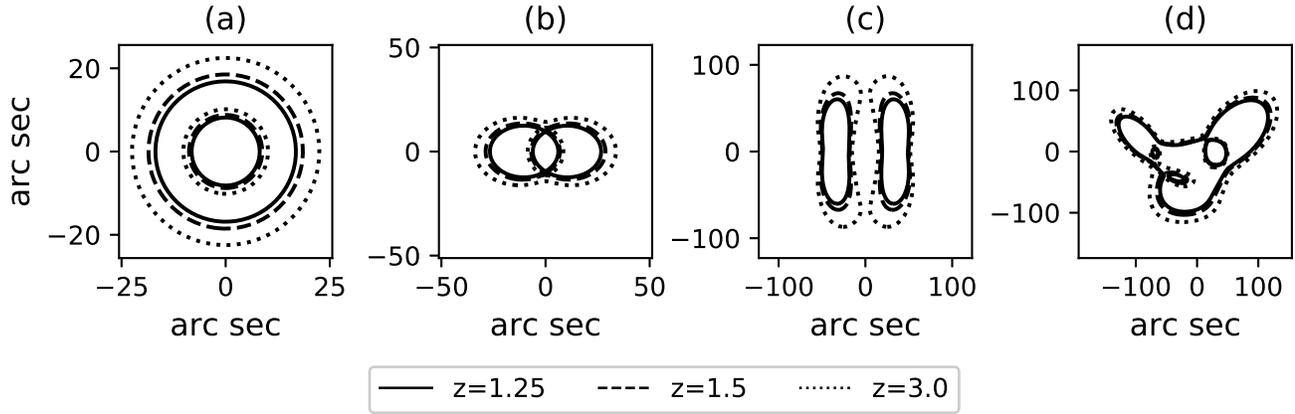}
    \caption{Critical curves generated by the four lensing models at the fiducial source redshifts of $1.25, 1.5$ and $3.0$.}
    \label{fig:critical}
\end{figure*}
\begin{figure*}
    \includegraphics[width=\textwidth]{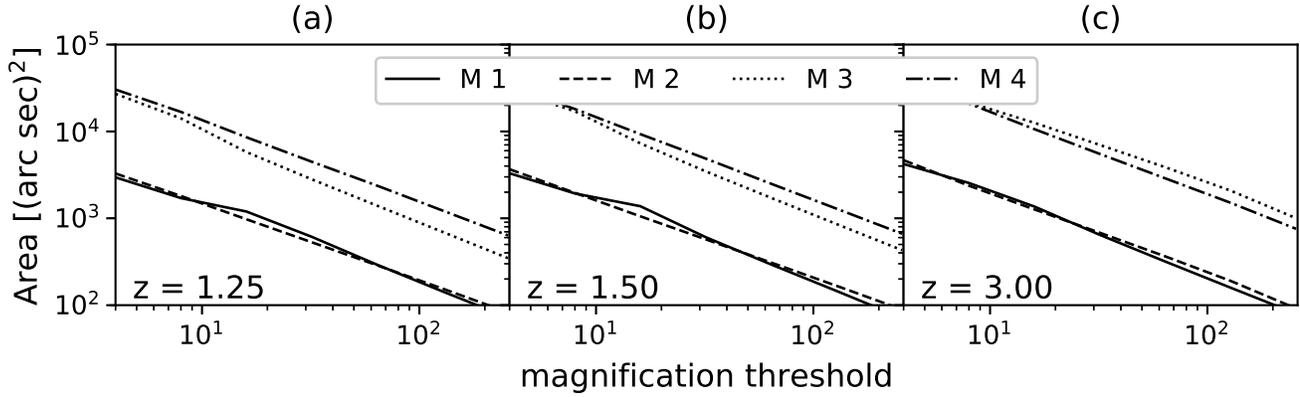}
    \caption{Area in arcsec$^2$ with magnification over a certain threshold is plotted. Each panel shows for all four potential with the panels (a), (b) and (c) corresponding to source redshift of $1.25, 1.5$ and $3.0$ respectively. The lens redshift is 0.3.}
    \label{fig:ThrArea}
\end{figure*}
\begin{table}
\begin{tabular}{l|r|c|l|c|c|c|c}
\hline
Models       &  $x_0$ &  $y_0$  & $\epsilon$ &  $\chi$  &  $\theta_s$  & $\theta_a$  & $\sigma_v$    \\ 
             &  ($'$) &  ($'$) &             & ($^{\circ}$)& ($'$)  & ($'$)  & (km sec$^{-1}$) \\
\hline  
\hline
M~1      & 0     & 0       & 0          & 0          & 10      & 500    & 1000           \\
\hline
M~2      & 0     & 0       & 0.1        & 0          & 10      & 500    & 1000           \\
\hline
M~3      & -35   & 0       & 0.25       & 90         & 20      & 500    & 1100           \\ 
             & 35    & 0       & 0.25       & 90         & 20      & 500    & 1100           \\
\hline
M~4      & -50   & 0       & 0.30       & 45         & 10      & 500    & 1000           \\
             & 50    & 0       & 0.2        & 135        & 10      & 500    & 1500           \\
\hline
\end{tabular}
\caption{Parameters of the PIEMD potential for the four lens models considered in this work.}
\label{tab:LensMod}
\end{table}

Strong lensing by galaxy clusters are studied very widely in literature \citep{2007ApJ...662..781R, 2008A&A...489...23L, 2010MNRAS.402L..44R, 2015MNRAS.452.1437J} where multiple images of lensed galaxies are used to reconstruct the gravitational potential of the lens. Many different approaches to model the projected lensing potential exists. \citet{1998ApJ...502..531B} use a Soften Power-law Elliptical Potential (SPEP) model to develop a computational algorithm for estimating the lensing potential. A semi analytical model of dark matter halo with different profiles like NFW \citep{2002ApJ...566..652L}, soften and singular  isothermal potential are discussed in \citet{2002ApJ...566..652L}.  
The basic idea behind the reconstruction of the lensing potential in most of the methods is to start with a parametric model of the lensing potential and reconstruct the source structure given the lensed images \citep{2012MNRAS.420..155K}. The parameter of the model is then tuned, mostly using an MCMC variant, to find the best approximation to the lensing potential \citep{2007NJPh....9..447J, 2015ApJ...813..102B}. A typical cluster potential has several components, one or two large dark matter halos along with smaller components arising from the gravitational potential of the cluster galaxies. The larger halos, however, are the most important elements for the overall lensing \citep{2007arXiv0710.5636E}.  \citet{1993ApJ...417..450K}  have used the Pseudo Isothermal Elliptical Mass Distribution (PIEMD) to model the individual cluster potentials. The PIEMD model is one of the widely used parametric models for the dark matter halo potentials of the galaxy clusters. 
 
 \begin{figure*}
    \includegraphics[width=\textwidth]{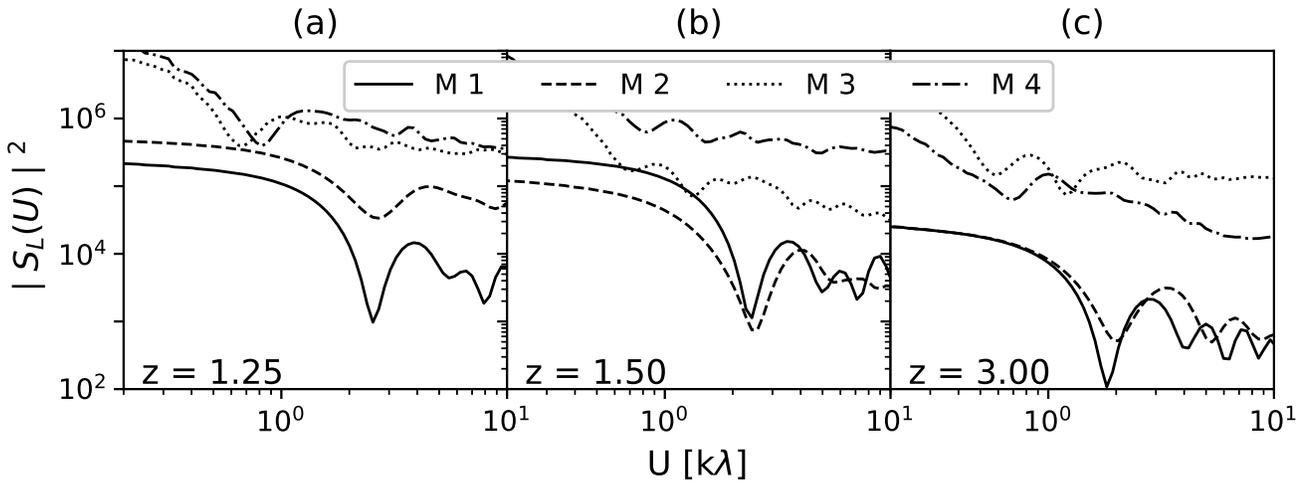}
    \caption{Modulus square of the azimuthally averaged lensing sampling function is plotted against baselines. Each panel shows for all four potential with the panels (a), (b) and (c) corresponding to source redshift of $1.25, 1.5$ and $3.0$ respectively. The lens redshift is 0.3.}
    \label{fig:SL}
\end{figure*}

 The projected potential for the PIEMD can be written as a function of the angular coordinates from the centre of the field of view of observations as
 \begin{eqnarray}
\psi(\theta_x, \theta_y) &=& 6\pi \frac{D_{ds}}{D_s} \frac{\theta_s + \theta_a}{\theta_s} \frac{\sigma_{v}^2}{c^2} \ f(\theta),  \\ \nonumber 
f(\theta) &=&   \sqrt{\theta_s^2+\theta^2}-\sqrt{\theta_a^2+\theta^2} \\ \nonumber
	 &+& \theta_a \ln \left( \theta_a+\sqrt{\theta_a^2+\theta^2} \right) -\theta_s \ln \left (\theta_s+\sqrt{\theta_s^2+\theta^2} \right),
\end{eqnarray}
where $D_{ds}$ and $D_s$ are the  distance between lens and source and the source and the observers respectively. The parameters $\theta_s$ and $\theta_a$ are the cut and the core radius, $\sigma_v$ gives the dark matter velocity dispersion to gravitationally support the halo. The angular variable $\theta$ depends on the angular coordinates $\theta_x$ and $\theta_y$ as 
\begin{equation}
\theta^2=\left[ \frac{ \theta_x\cos\chi +\theta_y\sin \chi }{1+\epsilon}\right] ^2+ \left [ \frac{-\theta_x\sin \chi+\theta_y\cos \chi}{1-\epsilon} \right] ^2,
\end{equation}
where $\epsilon$ is the ellipticity of the halo and $\chi$ gives the position angle of the ellipse. Each of the PIEMD component in the dark matter halo of the galaxy cluster is defined by seven parameters.

Various studies   of strong lensing by galaxy clusters in optical wavelengths have estimated the lensing potentials. They show that a typical lensing potential has several PIEMD components and estimate the seven parameters of each of these components. In this work, we consider four simple models of the cluster potentials based on PIEMD parametrization. Though these choices of parameters do not reflect any particular observed galaxy cluster, the parameters are chosen to lie within the known ranges of the observed galaxy clusters (see for reference \citet{2005MNRAS.359..417S, 2014ApJ...797...48J, 2018ApJ...859..159C}). Here is a brief description of the four models we choose here:

\begin{itemize}
\item {\bf M~1}: A single circular pseudo isothermal halo.
\item {\bf M~2}: A single pseudo isothermal elliptical halo. 
\item {\bf M~3}: Two component PIEMD cluster halo with the  same parameters except for the position of their centres.
\item {\bf M~4}: Two component PIEMD cluster halo with different parameters for the components.
\end{itemize}
The gravitational lens at a redshift $z_l$ only modifies the specific intensity from redshifts $z>z_l$. In this work, we consider all our lens models are at a redshift of $0.3$ and $z_l=1.0$. We first discuss the result with $z_l=0.3$ in the next section.

\subsection{Point spread function of the gravitational lens}
Given a projected lensing potential $\psi(\theta_x, \theta_y)$, the matrix $A_{ij}(\theta_x, \theta_y)$  can be estimated at each angular positions as 
\begin{equation}
 A_{ij}(\theta_x, \theta_y) = \delta_{ij}-\frac{\partial^2 \psi}{\partial \theta_i \, \partial \theta_j}, 
 \end{equation}
where $(i,j)$ corresponds to the combinations $(x,x), (x,y), (y,x), (y,y)$ etc, $\delta_{ij} = 1$, if $(i,j) = (x, x)$ or $(y,y)$ and $0$ otherwise. The lensing magnification function $\mu(\theta_x, \theta_y)$ can be calculated as the inverse modulus of the determinant of the matrix $A_{ij}$. \citet{10.1093/ptep/pty119} show that the lensing magnification function can be approximated as the point spread function of the gravitational lens when the dimension of the lensing potential is much larger than the wavelength of the lensed radiation and interference effects can be neglected. Hence, we may write, 
\begin{equation}
G_L(\theta_x, \theta_y) = \frac{1}{det \mid A_{ij} (\theta_x,\theta_y)\mid}.
\end{equation}
We generate a grid in the image plane to estimate the magnification function and hence the point spread function. We show the critical curve for a lens redshift of $0.3$ and three different source redshifts for our four lens models in  Figure~\ref{fig:critical}. The critical curves for the circular model are circular. As the lensing models get complicated the critical curves also show interesting features. 

The magnification function of the strong gravitational lens is expected to be unity away from the critical curves. Near to the critical curve the magnification increases. An important aspect of the lensing models is the area in the image plane over which the magnification function has a value above a certain threshold \citep{2001ApJ...557..421S}. We plot this area in arc sec $^2$ as a function of the  magnification threshold in Figure~\ref{fig:ThrArea}. In each panel, we show results from four models with different panels corresponding to different source redshifts. Lens redshift is kept fixed at $z_L=0.3$. Clearly, the single halo models have significantly lower magnifications compared to the double halo models. The area for a given model increases slowly with the redshifts. Within the two single and two double halo model we do not notice much difference in the effectiveness of the magnification reflected in these plots.

\subsection{Results}
\begin{figure*}
    \includegraphics[width=\textwidth]{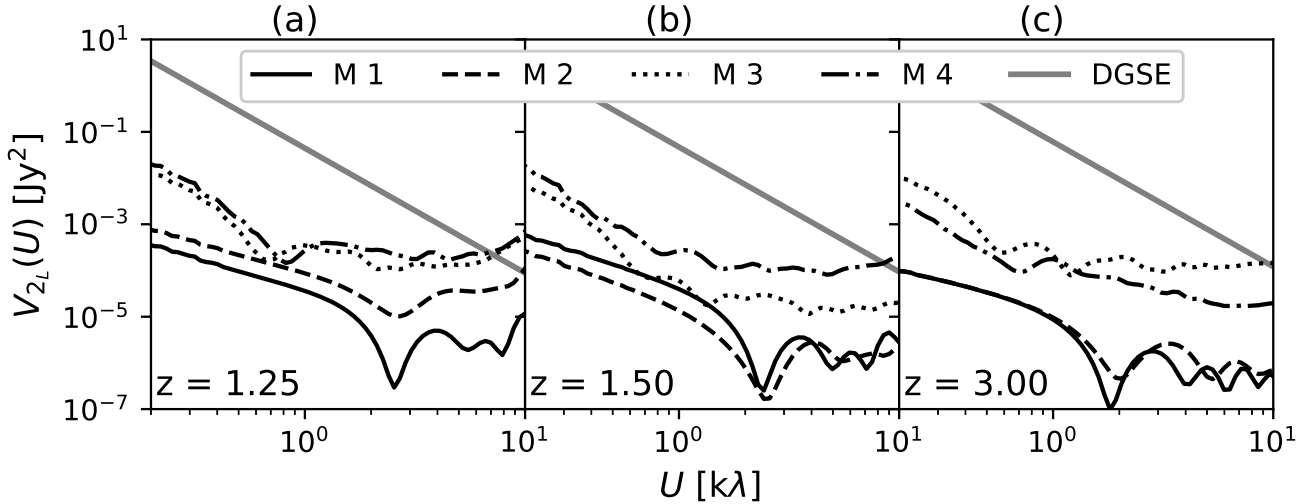}
    \caption{Visibility correlation in the presence of strong gravitational lensing is shown as a function of baseline. Each panel shows for all four potentials with the panels (a), (b) and (c) corresponding to source redshift of $1.25, 1.5$ and $3.0$ respectively. The grey line corresponds to the expected diffused galactic synchrotron radiation foreground (DGSE) from \citet{2016ApJ...818..139T}. The lens redshift is 0.3.}
    \label{fig:V2L}
\end{figure*}
\begin{figure*}
    \includegraphics[width=\textwidth]{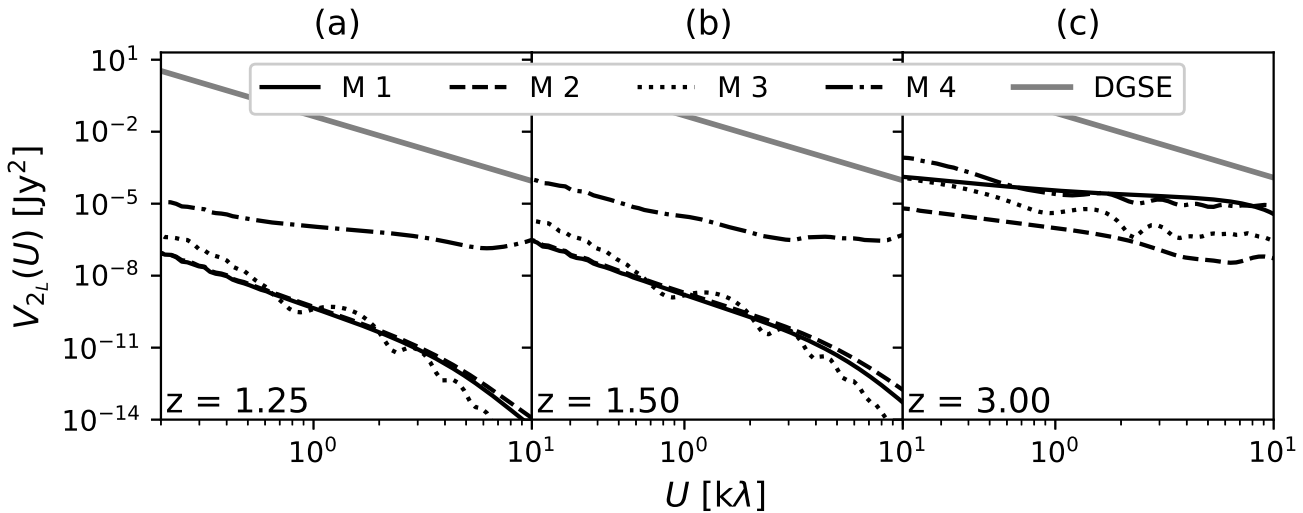}
    \caption{Visibility correlation in the presence of strong gravitational lensing is shown as a function of baseline. Each panel shows for all four potentials with the panels (a), (b) and (c) corresponding to source redshift of $1.25, 1.5$ and $3.0$ respectively. The grey line corresponds to the expected diffused galactic synchrotron radiation foreground (DGSE) from \citet{2016ApJ...818..139T}. The lens redshift is 1.0.}
    \label{fig:V2L1}
\end{figure*}

We calculate the Fourier transform of $G_L$ and estimate the modulus square of the azimuth averaged lensing sampling function $\mid S_L \mid^2$. Each panel of Figure~\ref{fig:SL} shows the variation of $\mid S_L \mid^2$ with baselines for all four lens models for $z_L= 0.3$. Note that the point spread function of the gravitational lens and hence the lensing sampling function depends on both the source and the lens redshifts. Three panels (a), (b) and (c) corresponds to the three different source redshifts. The sharp increase of the magnification function near to the critical curves results in oscillation in the sampling function and can be seen clearly in all models at larger baselines. The single halo models (M~1 and M~2) show the almost constant value of the $\mid S_L \mid^2$ at  baselines corresponding to the Einstein radius of each model and then a sharp decrease. At higher baselines, the single halo models have strong oscillatory features. The double halo models, in general, have higher values of $\mid S_L \mid^2$ compared to the single halo models. This is expected from the Figure~\ref{fig:ThrArea} where the same models show a significantly large portion of the area over a certain high magnification threshold. The oscillations in the double halo models are also subdominant because their corresponding magnification function has larger values at different places in the image planes. 

We use the eqn~(\ref{eq:V2L}) to estimate the lensed visibility correlation for our four lens models. The azimuthally averaged visibility correlation in the presence of the four lens models are shown in Figure~\ref{fig:V2L} for the lenses at a redshift of $0.3$. Different panels (a), (b) and (c) shows the results for three different redshifts $1.25$, $2.5$ and $3.0$ of the source 21-cm emission. The one halo models show the characteristics oscillations as in Figure~\ref{fig:SL} and have in general lower amplitude than the two halo models. The lensed visibility correlation for the two halo models is over $\sim 100$ $\mu$Jy$^2$ for all the baseline ranges plotted here. They show lesser oscillations as expected and explained earlier. 
A major challenge in detecting the redshifted \HI emission is the presence of strong foreground. The foreground contribution comes from the diffused galactic synchrotron emission (DGSE)  as well as the extragalactic compact sources \citep{2008MNRAS.389.1319J, 2016ApJ...818..139T, 2017arXiv170705335R}. Though there have been techniques to model and subtract the effect of compact source foregrounds in literature \citep{2017NewA...57...94C}, mitigating the DGSE is an outstanding challenge towards the detection of the redshifted \HI emission. An advantage of using the strong gravitational lensing to enhance the redshifted \HI signal is that the enhancement happens only for the \HI signal and not for the DGSE. The grey solid lines in the three panels in  Figure~\ref{fig:V2L} show the DGSE contributions at the frequencies corresponding to the redshifted \HI signal \citep{2016ApJ...818..139T}. We observe that with the two halo models, at least for a baseline range of investigation, the strong lensing enhances the \HI signal significantly and is expected to help in foreground removal.

All the results presented so far are for the lens redshift of $z_L=0.3$. To investigate the effect of lens redshift, we calculate the visibility correlations in the presence of strong lenses at a redshift of $1.0$ in Figure~\ref{fig:V2L1}. All the other parameters of the models including the lenses are kept same. Clearly, in this case, the enhancement of the visibility correlation due to strong lensing for all the model is rather less.

\section{Discussion and Conclusion}
In this work, we develop a formalism to estimate the power spectrum of the redshifted \HI 21-cm specific intensity using strong gravitational lensing. Using the visibility correlation to measure the power spectrum, we show how the lensing sampling function enhances the visibility correlation signal. We choose four simple fiducial models of gravitational lenses to demonstrate the effect of strong lensing and the expected visibility correlation signal for lens redshifts of $0.3$ and $1.0$ and for signal redshifts of $1.25$, $1.5$ and $3.0$. The fiducial lens models are chosen to represent the dark matter halo potential of galaxy clusters. We see that for the lenses at a redshift of $0.3$ the visibility correlation signal is significantly magnified at the smaller baselines and hence the larger angular scales. At large baselines, the lensed visibility correlation for single halo models show a sharp drop in signal and become oscillatory in nature.  Two double halo models, we use here show higher enhancement of the visibility correlation. Apparently, the required sensitivity for detecting the lensed \HI signal is about a few tens of mJy for visibilities (See Figure~\ref{fig:V2L}). The existing radio interferometers like the MeerKAT, the GMRT and future interferometers like the SKA mid  can achieve this sensitivity with only a few tens of hours of observations at lower baselines to carry out studies using this formalism. We show here that the effect of strong lensing is a multiplication of the observed unlensed visibilities by the  lensing sampling function $S_L$. The sampling function can be seen as a modification of the baseline distribution of the telescope, where it down weights the measured visibilities at certain baselines and up weights some others. The uncertainty in the power spectrum estimates highly depend on the baseline distribution  of a telescope. Hence, the sensitivity of a telescope and lens combination needs to be accessed together. We are presently working on a power spectrum estimator based on the basic idea presented in this paper and estimating  the feasibility of using this method with the known strong lenses for observations with the GMRT, the MeerKAT and the SKAmid. For the lenses at a redshift of $1.0$ the enhancement due to gravitational lensing is found to be less significant, limiting the effectiveness of this formalism for cluster lenses at higher redshifts.

A striking advantage of this approach is that the strong lensing by a lens at a redshift of $z_l$ enhances the signal from the redshifted 21-cm emission at a redshift of $z_s > z_l$, however, it does not enhance the diffused emission from the Galaxy, which acts as a foreground to the 21-cm signal.  The major challenge with the foreground signals like DGSE is that they are inherently several orders of magnitude larger \citep{2002ApJ...564..576D, 2003MNRAS.346..871O, 2008MNRAS.389.1319J, 2008MNRAS.385.2166A, 2011MNRAS.418.2584G, 2013MNRAS.433..639P} than the redshifted 21-cm signal and hence effects like residual foregrounds after the subtraction, residual gain errors add to bias in the estimators of the 21-cm power spectra \citep{2010ApJ...724..526D, 2020MNRAS.495.3683K}. We observe that for the best cases discussed in this work, that is for the  two double halo lens models at a redshift of $0.3$, the enhanced visibility correlation signal is still at only a few percent of  the expected galactic foreground. The lensing effect, however, enhances the \HI signal selectively and significantly. Hence, even though the lensed signal is less than the DGSE, the enhancement is still quite significant and would be quite important in subtraction of the continuum of DGSE signal.
 
Note that, the formalism developed here assumes that the gravitational potential of the lens galaxy clusters is already known. Gravitational lensing is well studied  with optical images and several techniques have been developed over the years to estimate the lensing potential.  At present, there exist several galaxy clusters with a model for lensing potentials (see for example \citet{2014MNRAS.444..268R}, \citet{2018ApJ...859..159C} etc.) . In ongoing work, we  are engaged in to survey all the existing cluster lens models in literature and find the best candidates to implement the formalism developed in this work. We note that the existing lensing models often have large uncertainties with larger uncertainties for the regions with high magnification \citep{2014MNRAS.444..268R}. This may limit the effectiveness of the present formalism. We expect with the advent of better algorithms to estimate the lensing potential the uncertainties in the models will decrease and the lensing models would be accurate enough to be used. 

The radio continuum signal from the galaxy clusters is expected to add to the lensed redshifted 21-cm radiation as an extra foreground element. At present, we have not estimated the contribution from the cluster continuum. Since the continuum signal from the cluster is not expected to vary over a small range of frequencies, we expect that a continuum subtraction based method can be used to mitigate the cluster continuum. We plan to investigate this in future work.

 In summary, this work introduces a new visibility based method to estimate
the redshifted 21-cm power spectrum. Based on the initial calculations with
model cluster halo potentials, the method itself looks promising. However, in
order to implement this method, we need to design an estimator of the power spectrum 
based on the lensed
visibility correlation and access its bias and variance. It is to be noted
that the variance of the estimator partly depends on the number
of independent samples one uses for one estimate. Since the lensing sampling
function enhances the visibility correlation at only limited regions in the
visibility plane, the method described here by itself may be limited by sample variance.
Furthermore, the sample variance depends on the actual lens model and the
angular scale or baseline in concern. We are investigating the practical
implementation of the lensed visibility correlation method and assessments
of the reliability of the estimates combining  the best known strong lens
candidates from optical studies. This results will be presented in detail in  
future work. 

\section*{Acknowledgement}
PD and UA acknowledge useful discussion with Tapamoy Guha Sarkar. UA acknowledge GATE fellowship for funding this work.
UA is thankful to Meera Nandakumar, Jais Kumar and Pavan Kumar Vishwakarma for useful discussions during this work. Authors thank the anonymous referee for providing very useful comments and pointing our mistakes in a calculation in the initial version of the paper. 

\section*{Data Availability}
No new data were generated or analysed in support of this research.
\newcommand{\newblock}{}
\bibliographystyle{mn2e}
\bibliography{reference}

\end{document}